\documentclass[onecolumn,authoryear]{els-mrw} 

\usepackage{amsmath,amssymb,amsfonts,amsthm,makeidx,graphicx}
\usepackage{txfonts}
\usepackage{helvet}

\usepackage{aas_macros}

\begin{document}

\chapter{Stellar Magnetism}\label{chapstelmag}

\author[1]{V\'{e}ronique Petit}%
\author[2,3]{Mary Elizabeth Oksala}%

\address[1]{\orgname{University of Delaware}, \orgname{Dept. of Physics \& Astronomy}, \orgdiv{Bartol Research Institute}, \orgaddress{Newark, DE, USA 19716}}
\address[2]{\orgname{California Lutheran University}, \orgdiv{Physics Department}, \orgaddress{Thousand Oaks, CA, USA 91360}}
\address[3]{\orgname{Paris Observatory}, \orgdiv{LESIA}, \orgaddress{Meudon, France 92195}}

\articletag{Chapter Article tagline: update of previous edition,, reprint..}

\maketitle

\begin{abstract}[Abstract]
Stellar Magnetism affects all spectral types and exists and varies throughout the evolution of stars.  Magnetic fields can affect not only the interior of stars, but also their circumstellar environments. 
In this chapter, we concentrate on the magnetic fields that can be measured at the surface of stars through the influence of the Zeeman effect on their spectra. We provide a brief introduction to the Zeeman effect and its associated light polarization. We discuss the state-of-the-art spectropolarimetric techniques that are used to detect, measure, and characterize surface stellar magnetic fields. 
We then present one of the most outstanding problem in stellar physics, mainly the origins of magnetic fields in massive OBA stars. We describe our current knowledge of the properties of known magnetic massive stars, such as their incidence, field distribution, topology, etc., and offer an outlook on the impact of stellar evolution on stellar magnetism.
\end{abstract}

\noindent\textbf{Keywords}:  Stellar magnetic fields; Massive Stars; Spectropolarimetry; Stellar Evolution; Stellar Atmospheres

\begin{glossary}[Glossary]
\term{Fossil magnetic field} a stellar magnetic field that is not generated in a contemporaneous dynamo process, but formed in some event, e.g. in stellar birth with the ISM or in a merger scenario\\
\term{Rotational Broadening}  a Doppler shift between the local spectrum from the advancing side and the receding side \\
\term{Spectropolarimetry} observational tool to understand how polarization varies with wavelength, combining spectroscopy and polarimetry  \\
\term{Stokes parameters} mathematical description of the polarization state of light   \\
\term{Zeeman pattern} resulting spectral line shape due to splitting of lines in the presence of a magnetic field \\
\term{Zero(Terminal)-age main sequence} Position of the main sequence on the HRD at the start(end) of hydrogen core-burning\\

\end{glossary}

\begin{glossary}[Nomenclature]
\begin{tabular}{@{}lp{34pc}@{}}
BinaMIcS & Binarity and Magnetic Interactions in various classes of Stars \\
BOB & B-fields in OB Stars \\
$B_{\ell}$ & surface-averaged longitudinal (line-of-sight) magnetic field; sometimes denoted $B_z$ \\
FAP & False alarm probability\\
ISM & Interstellar medium\\
LIFE & Large Impact of magnetic Fields on the Evolution\\
LSD &Least-Square Deconvolution\\
MHD & Magnetohydrodynamics \\
MiMeS & Magnetism in Massive Stars\\
SNR & Signal-to-noise ratio \\
ZDI & Zeeman Doppler Imaging\\
HRD & Hertzsprung–Russell diagram\\
MS & Main Sequence\\
ZAMS & Zero-age Main Sequence\\
TAMS & Terminal-age Main Sequence\\
\end{tabular}
\end{glossary}

\begin{BoxTypeA}[chap1:box1]{Learning objectives}

\begin{itemize}
\item The basic principles and application of the Zeeman effect and polarization in the magnetic fields of stars.
\item The process of using spectropolarimetry to detect magnetic fields in stars and determine field properties.
\item The general characteristics and channels of formation of the  magnetic fields found in OBA main sequence stars.
\item The statistical properties of OBA stars that are on the main sequence or prior to the main sequence.
\item The current understanding of changes that occur for evolved magnetic massive stars.
\end{itemize}

\end{BoxTypeA}

\section{Introduction}

Magnetic fields are a ubiquitous characteristic of space plasmas, and stars are no exception. 
Electric charges can interact with already-present magnetic fields, and generate magnetic fields of their own through their movements. 
The strength of magnetic fields in astrophysical environments range from a few micro-Gauss in the Interstellar Medium (ISM) to $10^{15}$ Gauss in extreme neutron stars. As a point of comparison, the magnetic field of the Earth is about 0.5 G, and that of an MRI machine is about 30 kG. 

Magnetic fields can be found in various stellar environments. Our own sun generates its own magnetic field by a dynamo effect in its outer layers. The overall surface magnetic field is not that large, only a few G. However magnetic spots have localized and tangled magnetic fields that can reach a few kG. These magnetic fields are very dynamic, as solar flares and the solar cycle nicely demonstrate. 
Most stars that are similar to our sun or lower mass also have magnetic fields that are generated by dynamos. 
For more massive stars, like the OBA-type stars, dynamo magnetic fields are expected to be produced deep in their cores, but not at the surface. Therefore, we should not be able to measure such core-produced magnetic fields at their surface. 

Because the main method to measure magnetic fields in stars relies on the light that comes from their surface, it is very difficult to probe the interior of stars unless they pulsate, which enables seismic studies of their interior properties. Therefore the large majority of stellar measurements of stellar magnetic fields are for surface magnetic fields, as will be detailed in Sect.~\ref{Bfield:sec:sec1}. 

It was rather a surprise then when magnetic fields were discovered at the surface of a subset of OBA-type stars. These magnetic fields were also very different from that of solar-type stars. The origin of these magnetic fields is still an important unresolved question in stellar physics. Some hypotheses exist (described in Sect.~\ref{Bfield:sec:origin}), however, they each have a hard time explaining all of the intriguing observed characteristics of magnetic massive stars (presented in Sect.~\ref{Bfield:sec:incidence} through \ref{Bfield:sec:evolve}). 

The presence of magnetic fields will have an important impact on the stellar structure and the star’s subsequent evolution. This said, they will not act alone, as other ingredients such as stellar winds, pulsations, binarity, etc. also play an important role. The interplay between all of these effects might seem daunting, but they also offer some unprecedented opportunities, as they often provide important complementary information.

\section{Detection of magnetic fields in stellar atmospheres}\label{Bfield:sec:sec1}

\subsection{Zeeman effect on atoms}

In 1896, Pieter Zeeman observed that in the presence of a magnet, sodium lines appeared to broaden.  This broadening is now called the Zeeman effect and with quantum mechanics, we can better understand why this broadening occurs.   In the presence of an external static magnetic ($B$) field, the potential energy levels available to an electron are modified according to $U = m \mu_b B$, where $m$ is the magnetic orbital quantum number and $\mu_b$ is the Bohr Magneton.  This modification of energy levels creates a splitting, and an electron then has multiple options to change energy levels, according to the rules that dictate allowed and forbidden transitions.  Therefore, this splitting creates multiple spectral lines, instead of a single one (see Fig. \ref{Bfield:fig:Zeeman}).  The measurement of the line splitting can give an estimate of the magnetic field strength, $B$.

\subsection{Polarization due to the Zeeman effect}

Polarization is a property of any transverse wave that describes the direction of the wave oscillations.  In electromagnetic waves, the direction is dictated by the electric field component.  Electromagnetic waves from stars are generally unpolarized, meaning that the light is emitted with a variety of polarizations.  Circularly polarized light is typically produced when light from a star interacts with a magnetic field.  Linear polarization can occur when stellar light scatters from a disk, as well as light interacts with a magnetic field.

Polarized light can be characterized by four values called the \textit{Stokes parameters} that describe the state of polarization of the light.  Stokes I corresponds to the total intensity of the radiation from the star or the typical intensity recorded in spectroscopy.  Stokes Q gives the percentage of light that is either horizontal- or vertical-linearly polarized light.  Stokes U gives the percentage of light that is linearly polarized at 45 degrees.  Finally, Stokes V gives the percentage of light that is either left- or right-handed circularly polarized light.  For stellar magnetism, we are looking to find a non-zero value for the Stokes V signal.

\begin{figure}[t]
\centering
\includegraphics[width=.45\textwidth]{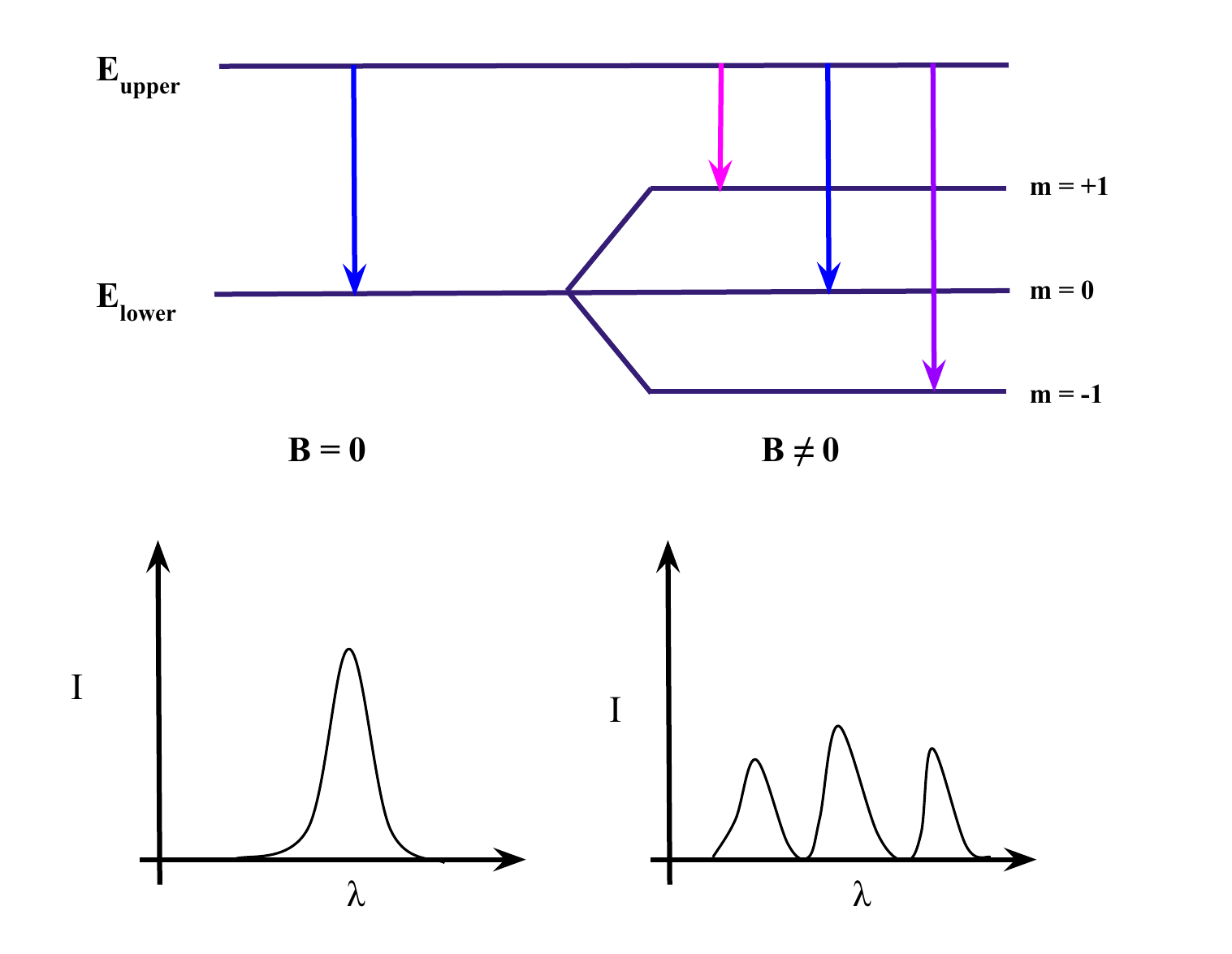}
\includegraphics[width=.45\textwidth]{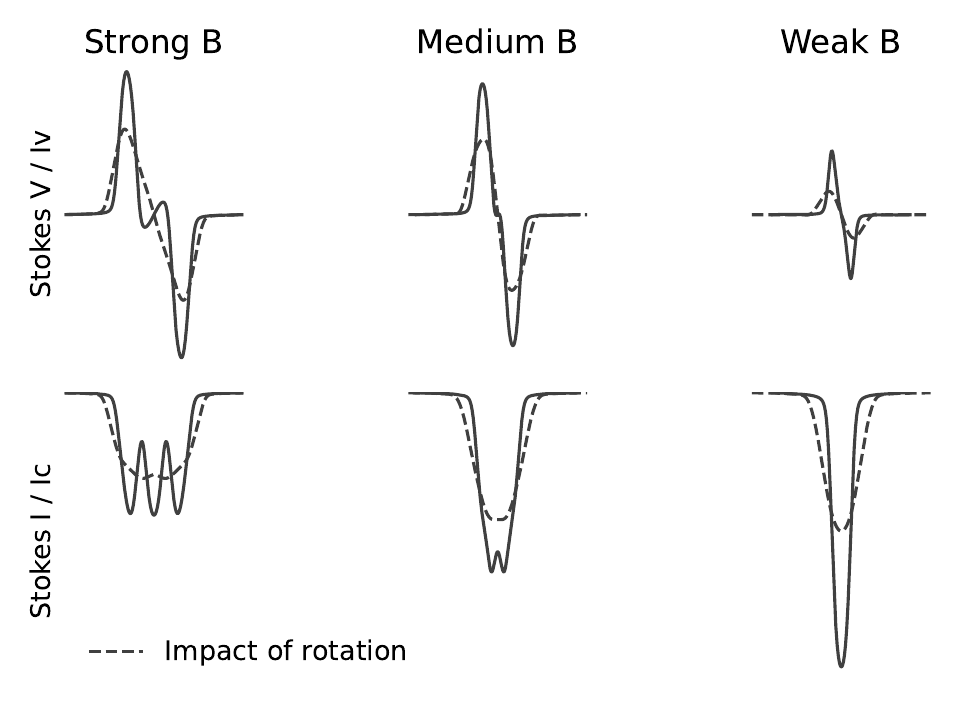}

\caption{\textit{Left:} Zeeman splitting of energy levels due to the presence of a magnetic field and the resulting intensity line profiles. On the left side is the case where $B=0$.  The right side shows the case where $B\ne0$. Here, we illustrate a so-called \textit{triplet} (where only one of the level splits into 3 sub-levels), but the numbers of sub-levels depends on the specific quantum numbers associated with the energy levels. The resulting set of transitions is called a Zeeman \textit{pattern}, and they can be quite varied \cite[see e.g.][]{2003isp..book.....D}.  Adapted from Figure 2 in \citet{2015IAUS..307..311L}.
\textit{Right:} Illustration of the impact of the dipolar field strength and stellar rotation on the Stokes I and V line profiles. Calculations made with the Unno-Rachkovsky solution \citep[see][for an overview]{2003isp..book.....D} for a triplet with $\beta=i=45^\circ$ and $\phi=0$. 
}
\label{Bfield:fig:Zeeman}
\end{figure}

In addition to the line splitting in the Zeeman effect, the multiple components of the transition will have different polarization states. This magnetic field effect can be broken down into two components.  The part of the field that is along the line-of-sight, often called the \textit{longitudinal} field, will polarize the transitions with $\delta m =\pm 1$ (the \textit{sigma} components) with orthogonal circular polarization states. The Stokes V parameter is the subtraction of these two states, and if there is a magnetic field present, we will observe a net change in Stokes V across a spectral line profile, even if the Zeeman splitting is not visible (as illustrated in Fig. \ref{Bfield:fig:Zeeman}). However, the circular polarization produced is not only dependent on the field strength, but also its orientation with respect to the observer.  The part of the field that is perpendicular to the line-of-sight, or the \textit{transverse} field component, will produce linear polarization along and perpendicular to the field axis. There will therefore be a net change in the Stokes parameter Q and U across the line of sight. The amount of polarization in each parameter depends on the orientation of the transverse field on the plane of the sky. It is important to note though that for the same field strength in the longitudinal and transverse directions, the linear polarization produced will always be an order of magnitude lower than the circular polarization. 

Circular polarization observations are therefore a really good choice in order to detect magnetic fields.  However, it is important to remember that the light coming from a spatially unresolved star is in fact the combination of the light produced at each point on the visible stellar surface, which can possess different local field strengths and orientations.  Because polarization produced by a certain point is sensitive to the orientation of the local field vector, some profile cancellation can occur, making the polarization more difficult to detect, and more difficult to interpret. The rotation of the star is also important to consider, as it introduces a Doppler shift between the local spectrum from the advancing side and the receding side. Because of this \textit{rotational broadening} in the Stokes V profile, the contribution to the polarization coming from two points of the star might not cancel out perfectly, and polarization can be seen even if the net field component in the longitudinal direction is zero.  An example of this can be seen in the middle plot in Fig. \ref{Bfield:fig:lsd}. \citet{2015IAUS..307..311L} is a good reference for a more detailed, but accessible description of the method of spectropolarimetry. 

\subsection{Spectropolarimeters}
To measure the polarization from a star requires a spectropolarimeter, which is an instrument that can record both the intensity or Stokes I and one or more of the Stokes QUV parameters as a function of wavelength or spectrum.  For stellar magnetism, the Stokes V spectrum is used to diagnose the presence of a field and multiple measurements throughout a star's rotation can be used to study the \textit{magnetometry} (the structure and geometry of the magnetic field).

A typical spectropolarimeter is placed into the incoming light beam of the telescope before the light is able to enter the spectrograph.  The instrument contains a series of optical components, including a prism to split the beam into its two orthogonal  polarization states so that the Stokes I spectrum and another Stokes parameter spectrum can be recorded simultaneously.  The instrument also includes both half and quarter wave plates that are able to rotate, so that all three Stokes polarization parameters can be recorded without having to replace any optical components.  In order to obtain a circular polarization or Stokes V spectrum, 4 individual observations must be made with the quarter and half wave plates at different angles that when combined they correspond to the mathematical representation for Stokes V.  Data from four exposures are also combined in two other ways to obtain what are referred to as the \textit{Null} spectra, as they do not combine to show any polarization, but instead allow the user to see if there are any optical residual signatures from the telescope or optical components within the instrument masking itself as real polarization signatures.

\begin{BoxTypeA}[]{Examples of instruments used to study stellar magnetism}
    
\begin{itemize}
\item{FORS2 (R $\sim$ 2000) spectropolarimeter: optical low-resolution instrument installed on one of the 8-m Unit Telescopes (UT1) of the European Southern Observatory (ESO)’s  Very Large Telescope (VLT) at the Paranal Observatory in Chile.  }
\item{ESPaDOnS (R = 65,000) spectropolarimeter: optical high-resolution instrument used on the 3-m Canada-France-Hawaii Telescope at the Mauna Kea Observatory in the USA.   This is a twin of the Narval instrument at the 2.2-m Bernard Lyot Telescope at the Pic du Midi Observatory in France.}
\item{HARPSpol (R = 105,000) spectropolarimeter: optical high-resolution instrument used on the 3.6-m ESO telescope at the La Silla Observatory in Chile.  }
\item{SPIRou (R = 75,000) spectropolarimeter:  near-infrared high-resolution instrument used on the 3-m Canada-France-Hawaii Telescope at the Mauna Kea Observatory in the USA.    }
\end{itemize}
\end{BoxTypeA}

\subsection{Measuring magnetic fields from spectropolarimetric data}
\label{Bfield:sec:measure}

To study magnetism in stars, we employ two types of techniques.  High-resolution spectra provide detailed information of the stellar photospheric structure, revealing any small-scale variability.  If a star is slowly rotating and has a strong magnetic field, the Zeeman splitting may be measured in individual lines, giving an estimation of the surface-averaged magnetic field strength.  However, the typical spectral separation will only be about 1-2 km s$^{-1}$ per kilogauss in the optical.  So, unless the field is quite strong, the splitting will be less than the typical Doppler broadening of $\sim$ 5-10 km s$^{-1}$. We may also use another additional type of observation called spectropolarimetry, which records spectroscopy and polarimetry simultaneously. Polarimetry provides a characterization of the magnetic structure of the star, including field strength, configuration, and geometry.  Together, these two observed techniques produce a comprehensive data set to analyze and model the physical parameters of the star.  Historically spectropolarimetry has been conducted using optical light, but infrared detectors are being used to study new types of objects, such as with SPIRou.

Spectropolarimetry can be used in two different methods, depending on the resolution of the instrument.  For low-resolution optical spectropolarimeters, such as the FORS2 instrument on the VLT (and its predecessor FORS1), surface-averaged longitudinal magnetic field measurements, $B_z$, are obtained using the relationship
$$
\frac{V}{I} = -g_\mathrm{eff} C_Z \lambda^2 \frac{1}{I}\frac{dI}{d \lambda} \langle B_z\rangle,
$$
where $B_z$ is the surface-averaged longitudinal magnetic field, $g_\mathrm{eff}$ is the effective Land\'e factor, $\lambda$ is the average wavelength, and $C_Z = \frac{e}{4\pi m_e c^2}$, where $m_e$ is the mass of an electron, and $e$ is the charge of an electron \citep[for a basic review of this technique see, e.g.,][]{2015A&A...583A.115B}.

\begin{figure}[t]
\centering
\includegraphics[width=\textwidth]{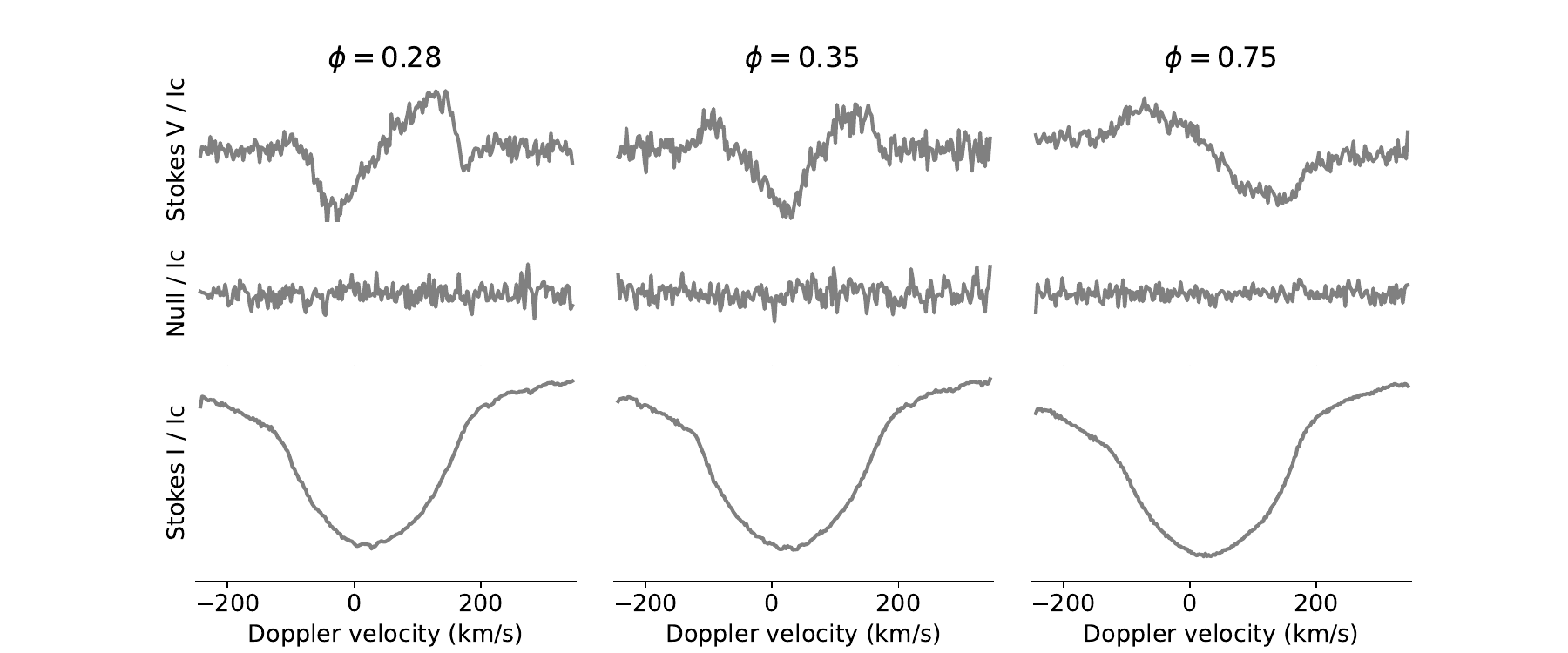}
\caption{LSD Stokes I (bottom), null N (middle), and Stokes V (top)
profiles for three different rotational phases of the proto-typical, He-strong Bp star $\sigma$ Ori E. For visualization
purposes, the null and Stokes V have been shifted upwards from 0.0. The profile at rotational phase 0.35 shows an example of a case where the longitudinal field is close to zero, but the Stokes V signature is sill visible because of rotational broadening. The change or flip in polarity is viewed by comparing the Stokes V profile shapes at phase 0.28 to phase 0.75. The LSD profiles used in this figure are from \citet{2012MNRAS.419..959O}.}
\label{Bfield:fig:lsd}
\end{figure}

For higher resolution optical or infrared spectropolarimeters, such as ESPaDOnS or SPIRou on CFHT, we can use the Stokes V data in individual lines to directly measure the surface-averaged longitudinal magnetic field strength ($B_{\ell}$, equivalent to $B_z$), if the field strength is large enough.  However, a multi-line technique is often used to identify smaller field strengths and increase the signal-to-noise ratio (SNR), called \textit{Least Squares Deconvolution} \citep[LSD,][]{1997MNRAS.291..658D}, which uses a list of spectral lines to make an effective average of the information contained in all of the lines simultaneously to produce an average Stokes I, null and Stokes V profile (see Fig. \ref{Bfield:fig:lsd}). This method assumes that the included lines have relatively similar \textit{shapes} with amplitudes dictated by the characteristics of the individual lines (depth, magnetic sensitivity, etc) -- therefore hydrogen lines are typically avoided.

Either using the individual signatures in lines or the LSD average profiles, $B_{\ell}$ can be derived from the Stokes V profile by computing the first moment of the Stokes V 
$$
B_{\ell} = -2.14\times 10^{11} \frac{\int v V(v)dv}{\lambda g c \int[I_c-I(v)]dv} ~[\mathrm{G}],
$$
where $g$ is the Land\'{e} factor (magnetic sensitivity) of the line or the average Land\'{e} factor if using the LSD line profile, $\lambda$ is the central wavelength (in nanometers) or the LSD average wavelength, and $c$, the speed of light, should have units of km s$^{-1}$ \citep{1997MNRAS.291..658D,2000MNRAS.313..851W}.  The integral is over velocity space, where the center of the line profile corresponds to $v=0$.  The range of integration should typically be set so that it covers the entire spectral line. The value obtained for the longitudinal magnetic field often shows a variability related to rotation modulation of our view of the magnetic field geometry.  To use this variability to estimate the dipolar field strength and the overall geometry of the system, i.e. the incidence angle, $i$, and the obliquity angle, $\beta$, we can employ the Preston Method, which is described in the boxed section below.

A statistical test called the False Alarm Probability (FAP) provides an objective test as to whether a magnetic field is detected for a star.  The FAP is computed using the $\chi^2$ probability function associated with how much the Stokes V (and N) signal deviated from zero \citep[see][for more specific details]{1992A&A...265..669D}.  A signal is definitely detected (DD) in either V or N when the associated probability is larger than 99.999\% or when the FAP is smaller than $10^{-5}$.  A marginal detection (MD) corresponds to a FAP that is between $10^{-5}$ and $10^{-3}$.  A FAP below $10^{-3}$ is considered a non-detection \citep[ND;][]{1997MNRAS.291..658D}.

\begin{BoxTypeA}[Bfield:Preston]{The Preston method \citep{1950MNRAS.110..395S,1967ApJ...150..547P}}

\noindent The change in $B_\ell$ as the star rotates can be predicted under some simplifying assumptions:
\begin{enumerate}
\item{The magnetic field is a dipole centered with the center of the star.}
\begin{itemize}
\item{The magnetic field can be tilted by an angle $\beta$ with respect to the axis of rotation (the \textit{obliquity} angle).}
\item{The axis of rotation can be tilted by an angle $i$ with respect to our line-of-sight to the star (the \textit{inclination} angle)}
\end{itemize}
\item{The temperature structure in the photosphere of the star is such that we can assume that the source function of the radiation is linear with optical depth, which results in a so-called \textit{linear Limb-Darkening}.}
\end{enumerate}
In this case, the relation between $B_\ell$ and the strength of the dipolar field $B_p$ is:
$$
B_\ell = \frac{B_p}{20}\frac{15+u}{3-u}(\cos\beta\cos i+ \sin\beta\sin i\cos\phi),
$$
where $\phi$ is the rotational phase with $\phi=0$ corresponding to when a magnetic pole is the closest to our line-of-sight, and $u$ is the limb-darkening coefficient defined as $u = 1-I_\mathrm{limb}/I_\mathrm{center}$ when $I$ is the intensity of the light at the center and limb of the stellar disk. 

If we define a quantity $r$ as being the ratio of the minimum value to the maximum value of $B_\ell$ ($r=B_{\ell,\mathrm{min}}/B_{\ell,\mathrm{max}})$ and noting that these values of  $B_\ell$  will occur when $\phi$ is $0$ and $\pi$, respectively, we can obtain a constraint between $i$ and $\beta$ that is independent of the polar strength or the limb-darkening coefficient:
$$
\tan\beta = \left(\frac{1-r}{1+r}\right)\frac{1}{\tan i} 
$$
From a measured variation of $B_\ell$ (datapoints in figure below), we can employ e.g. sinusoidal fits (curves in left panel) to determine $B_{\ell,\mathrm{max}}$ and $B_{\ell,\mathrm{min}}$ (dashed lines), compute $r$, and obtained the constraints showed by the curves in the right panel. 
If one has a separate measurement of the inclination angle (for example from the combination of the rotation period, the rotational line broadening, and the star’s radius), the obliquity angle can then be constrained. 

\begin{centering}
\includegraphics[width=0.7\textwidth]{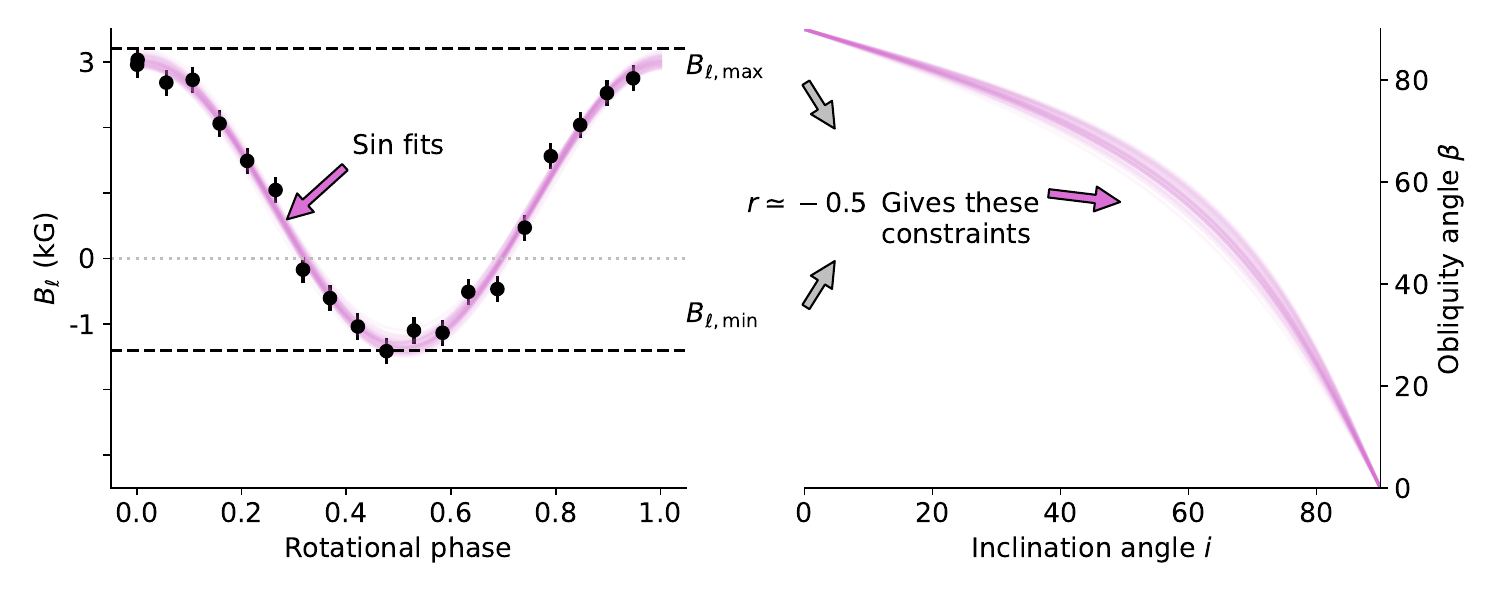}
\end{centering}

\end{BoxTypeA}

While there are many computational programs and tools to analyze spectropolarimetric observations of stars using various versions of Fortran, IDL, IRAF, etc., there has been a recent effort to create a centralized suite for analysis entirely in python. SpecpolFlow (Folsom et al., in prep) is a pythonic workflow that is a single pipeline for the analysis of spectropolarimetric observations of astronomical sources.  The SpecpolFlow package is available at github.com/folsomcp/specpolFlow.  In addition to building all of the tools necessary to take data directly from the telescope through to a full analysis with LSD, the team has also built a number of tutorials to ensure the tools are accessible to even novice users of spectropolarimetry. 

If the Stokes V profile is rotationally modulated and one is able to obtain a time-series of spectropolarimetry sufficiently covering the rotation of a magnetic star, the technique of Zeeman Doppler imaging (ZDI) can be used to recover information about the magnetic topology and more accurately determine the geometry of the system (inclination, obliquity). ZDI combines the Stokes QUV lines from spectropolarimetry and a technique that uses Doppler broadening to  map inhomogeneous structures on stellar surfaces called Doppler imaging.  It was first proposed to be applied for magnetic fields by \citet{1989A&A...225..456S}  and first implemented by e.g., \citet[][]{1991A&A...250..463B}, inverting the Zeeman signatures onto a vector magnetic field surface distribution.  Since then many implementations of this basic premise have led to mapping of magnetic structures.  Current methods break down the magnetic structure of the poloidal or toroidal components of the field using spherical harmonic expansion \citep[e.g.,][]{2006MNRAS.370..629D,2015MNRAS.453.2163S}.  This method of ZDI works for all types of magnetic stars, and can even use the LSD Stokes V profiles as input.

\section{The intriguing origin of magnetic fields in hot stars}

\subsection{The definition of fossil fields}

The magnetic fields of solar-type and low-mass stars are generated contemporaneously by a dynamo mechanism. 
This dynamo operates in the large convection zones present in these stars’ outer envelopes.
The structure of a star is set by the temperature structure that determines where convection or radiation is more efficient to move energy from the core to the surface.
For stars on the main sequence with masses larger than a few times the mass of our sun, the transport of energy in their outer layers can be done through radiation and no significant convection is expected in their outer-layers \citep[see][for details]{2012sse..book.....K}.

It was therefore surprising that surface magnetic fields on a star other than our own Sun were first detected in an intermediate-mass star \citep[78 Vir, ][]{1947ApJ...105..105B}.
Now, it has been established that about 10\% of stars with \textit{radiative} envelopes (intermediate stars and massive stars on the main sequence) have surface magnetic fields.
Given that these fields are not from a solar-type dynamo, they have been dubbed \textit{fossil fields}. The term originated from initial ideas that these fields might be caused by the concentration of ISM field lines during star formation \citep[e.g.][]{1965IAUS...22..420M}. Nowadays, the term \textit{fossil} refers to any magnetic field that is not contemporaneously generated by a dynamo — they are a leftover from a “thing” that occurred in the star’s past, may it be an evolution phase, an event like a binary merger, or a mechanism during star formation.
This said, it is important to note that some publications use “fossil origin scenario” to only mean the scenarios in which the fields originate from star formation, and do not refer to a magnetic origin from e.g. a binary interaction (that occurred in the past but after the star was born) as “fossil” --  therefore some caution is needed. 

\subsection{Possible channels of formation}
\label{Bfield:sec:origin}

\subsubsection{ISM}

Magnetism is quite ubiquitous in all stages of star formation, from molecular filament to protostars \citep[see][]{2023ASPC..534..193P}. It would therefore not be surprising to find fossil stellar magnetic fields at the surface of main sequence stars. If anything, it could be viewed as rather strange that only 10\% of main sequence massive stars end out being magnetic. 
The basic idea of the ISM fossil field hypothesis is that the ISM magnetic field can be pulled in and enhanced during the gravitational collapse. Some of the magnetic flux will inevitably be lost during this process due to various dissipation mechanisms, and only stars formed in areas of large enough initial fields would end up being magnetic (e.g., Mestel 1999). 
Some studies have shown that a magnetic field might inhibit the fragmentation of star-forming clouds \citep[e.g.][]{2008MNRAS.385.1820P}. There might therefore be a link between magnetism and mass, and/or between magnetism and binarity.

\subsubsection{Mergers}

Another interesting proposed channel for magnetism could be through binary interaction, where the magnetic fields are a by-product of a mass transfer or a merger event. 
About 30\% of massive stars are expected to interact in their lifetime, and about 10\% of them are even expected to merge into a single object \citep{2014ApJ...782....7D}. 
During such an event, shear should give rise to magnetic fields according to MHD simulations \citep{2019Natur.574..211S}. It is the general consensus that merger products should appear “rejuvenated” (that is, appear younger than they really are) and fast rotating. However, recent theoretical studies of mergers incorporating magnetic fields \citep{2020MNRAS.495.2796S} have shown that fast rotation might not always be the typical outcome. 

\subsubsection{Subsurface convection}

Main sequence massive O and B stars do have convection zones in their cores. There is both theoretical and observational (through asteroseismology) evidence that magnetic fields are generated in these convection zones \citep{2016arXiv160303659A,2009ApJ...705.1000F,2022MNRAS.512L..16L}. It is however understood that i) these dynamo-generated, core magnetic fields cannot reach the surface \citep{2004MNRAS.348..702M} and ii) even if they could, we would then expect \textit{all} stars with convection cores to have some degree of surface magnetic fields that correlate with the core properties — core dynamos therefore cannot readily explain the fossil field phenomenon.

Although we usually say that massive stars on the main sequence lack convection zones in their envelopes, we actually mean \textit{substantial} convection zones. Massive stars with radiative envelopes still host thin, inefficient convection zones that are caused by opacity peaks from the ionization of hydrogen, helium, or iron-peak elements. These convection zones might be able to operate a dynamo, however, the magnetic fields emerging from these studies are usually expected to be weak and small-scaled \citep[see][for more details]{2022ApJS..262...19J}.

\subsection{Incidence of magnetism in un-evolved OBA stars}
\label{Bfield:sec:incidence}

Magnetic A and late-B (from spectral class B5 and cooler) stars on the main sequence have been better studied than their more massive counterparts, for a few reasons.
\begin{enumerate}
\item They have more spectral lines from metals in their optical spectra, which means that the multi-line techniques (REF to earlier section) are more effective. \item Their spectral lines are narrower than those of hotter stars (less thermal, Stark, and rotational broadening), which means that the Stokes V signal has a higher amplitude for the same magnetic field value and therefore requires a smaller signal-to-noise ratio to detect.
\item They are more numerous in our local environment because the Initial Mass Function (see the Chapter on the IMF in this Encyclopedia) favors less massive stars. 
This helps with building enough statistics to enable general conclusions about their properties. 
\end{enumerate}
All of these properties make up for the fact that A and late-B stars are intrinsically less luminous than O and early-B stars. 

Another important aspect that has enabled better studies of magnetic A and late-B stars is that they are easier to find by indirect methods. The atmospheres of these stars are very stable, which means that the chemical settling of various elements can take place without being disturbed by turbulence, outflows, or convection (see the chapter on chemically peculiar stars in this Encyclopedia). 
Magnetic A and late-B stars develop a specific abundance peculiarity pattern that is designated \textit{ApBp}. There is a two-way relationship between magnetism and the ApBp classification: all strongly magnetic A and late-B stars have ApBp chemical peculiarity, and all stars with the ApBp chemical peculiarity have been found to be magnetic \citep[e.g.][]{2007A&A...475.1053A}. 
This means that magnetic A and late-B stars can be identified through spectroscopy only. 

Furthermore, the distribution of elements is often not uniform at the surface of ApBp stars. There are patches of high or low abundances of various elements that often correlate with the structure of the magnetic field. Different abundances change the local flux emerging from various parts of the visible stellar disk. If there is an asymmetry in these patches with respect to the rotation axis of the star, the combined flux coming from the visible stellar disk will change as the star rotates. This means that photometric variations can also be used to identify ApBp stars \citep[see e.g.][]{2023A&A...676A..55L}. 

For hotter O and early-B stars, there are no direct two-way relationships between magnetism and chemical peculiarly. Some stars do display:
\begin{itemize}
\item the Of?p star classification \citep{1972AJ.....77..312W}, which relates to the relative strength of C\textsc{iii} $\lambda\lambda$4647–4652\AA\ and the N\textsc{iii} $\lambda\lambda$4634–4642\AA\ lines (all Of?p stars have been found to be magnetic, but not all magnetic O-type stars have this spectral peculiarity). 
\item helium abundance peculiarities and He-line variations \citep[e.g.][]{1987ApJ...323..325B},
\item variable and peculiar emission from their circumstellar \textit{magnetospheres} \citep[see review by][for details]{2022hxga.book...46U}.
\end{itemize}
Periodicity is therefore a good way to identify O and early-B magnetic candidates, and some surveys have successfully exploited this to pre-select stars with good likelihood of being magnetic \citep[e.g.][]{2018MNRAS.478.2777B}. 
However, to compute the incidence of magnetic fields without selection-biases, it is necessary to confirm the presence of a magnetic field with spectropolarimetry for O and early-B stars.

\begin{BoxTypeA}[Bfield:rotation]{Interesting fact}

The rotation period of sun-like stars is easily determined by photometric variations due to magnetic spots. For magnetic massive stars, the photometric and spectral variations similarly can provide a rotational period as well. This period can sometimes be ambiguous, because the spectral and photometric diagnostics are not sensitive to the polarity of the magnetic field, which can sometime result in ambiguities between different period aliases. However, the magnetic fields of massive stars usually have a simple topology, and the variation of the longitudinal magnetic field measurements, which is sensitive to the field polarity, can typically lift these ambiguities. 

For the 90\% of massive stars that do not have a magnetic field, however, there are no good methods to identify rotation periods with good certainty. The line broadening gives us the equatorial velocity $v_\mathrm{eq}$ multiplied by the sin of the inclination of the rotational axis with respect to our line of sight $i$ ($v_\mathrm{eq}\sin i$). Fast rotating stars can be readily identified by their large $v_\mathrm{eq}\sin i$, but slow rotators are more difficult to identify as a fast rotating star could masquerade as a slowly rotating star if it has a small inclination angle $i$. Magnetic massive stars with long rotational period are therefore the perfect laboratory to e.g. study various types of non-rotational broadening such as macro-turbulence \citep{2013MNRAS.433.2497S}. 
\end{BoxTypeA}

\subsubsection{Incidence of main sequence magnetic ApBp and pre-main sequence Herbig AeBe stars }

The general incidence of magnetism in main sequence A and late-B stars is around 10\%. 
Fig. 17 of \citet{2019MNRAS.483.2300S} shows the distribution of incidence as a function of mass that was obtained from a 100-pc volume limited survey of ApBp stars. In their lower mass bin (1.4 - 1.6 $M_\odot$), the incidence is basically zero and rises to 10\% by $\sim 3M_\odot$. 
This means that the transition from ubiquitous solar-type dynamo magnetism to a 10\% of fossil field is not direct -- there is a gap in magnetism around spectral type early-F star \citep[see also Fig. 1 of][]{2020MNRAS.494.5682S}. 

The presence of magnetic fields in A and late-B stars does not seem to correlate with stellar characteristics other than the ApBp phenomenon, which is a surface effect. ApBp stars occupy the same regions of the HRD as normal A and late-B stars \citep[see Fig. 14 in][]{2019MNRAS.483.2300S}. However, the bulk of the ApBp star population seems to rotate slower than their non-magnetic counterparts \citep{1995ApJS...99..135A}, which is usually understood to be a consequence of magnetic braking. 

Interestingly, fossil magnetic fields with similar characteristics to that of ApBp stars are already present during the pre-main sequence phase for these intermediate-mass stars. In a spectropolarimetric survey of about 70 Herbig AeBe stars,  \citet{2013MNRAS.429.1027A,2013MNRAS.429.1001A} found that ~7\% of them host magnetic fields, all stronger than 300G.

\subsubsection{Incidence of main sequence magnetic O and early-B stars}

Similarly, for main sequence O and early-B stars, all large-scale surveys agree that the general incidence is a bit lower than 10\%. 
Unlike for ApBp stars, massive OB stars are not as numerous in our local space, therefore volume-limited surveys are impractical. Therefore, most large-scale surveys are instead magnitude-limited, and prone to numerous selection biases. 
To give a sense of numbers, the \textit{Magnetism in Massive Stars} (MiMeS) survey \citep{2016MNRAS.456....2W} has found 6 new magnetic stars in a sample of 97 O-type stars \citep{2017MNRAS.465.2432G}, and the \textit{B Field in OB Stars} (BOB) survey \citep{2014Msngr.157...27M} has found 4 new magnetic stars in their sample of 69 OB-type stars \citep{2017A&A...599A..66S}. 

Again, there is no evidence of correlation of the incidence of magnetic field with stellar parameters, other than a loose correspondence with surface effects such as helium abundance peculiarities, magnetospheric emission, and the Of?p phenomenon. Magnetic O and early-B stars seem to roughly occupy the same region of the HRD as their non-magnetic counterparts \citep[e.g. see Fig 10 from][]{2017MNRAS.465.2432G}. 
The rotation rates of magnetic massive OB stars are quite diverse. On one hand, most of the dozen or so known magnetic O-type stars have been found to be slow rotators \citep[see the list of periods in Table 1 of][]{2013MNRAS.429..398P}. Some periods are extremely long \citep[for example the O8f?p star HD 108 with a 55-yr rotational period,][]{2017MNRAS.468.3985S}. 
On the other hand, for magnetic early-B stars, some periods are measured in years \citep[e.g. the B0.5 star $\xi^1$ CMa with $P\sim30$ yr,][]{2021MNRAS.506.2296E}, whereas some are less than a day and nearly at critical velocity \citep[e.g. the B2 star HR 7355 with $P\simeq0.5$ d,][]{2010MNRAS.405L..51O}. 

\subsection{Distribution of magnetic field strengths and the magnetic desert on the main sequence}

\begin{figure}[t]
\centering
\includegraphics[width=.5\textwidth]{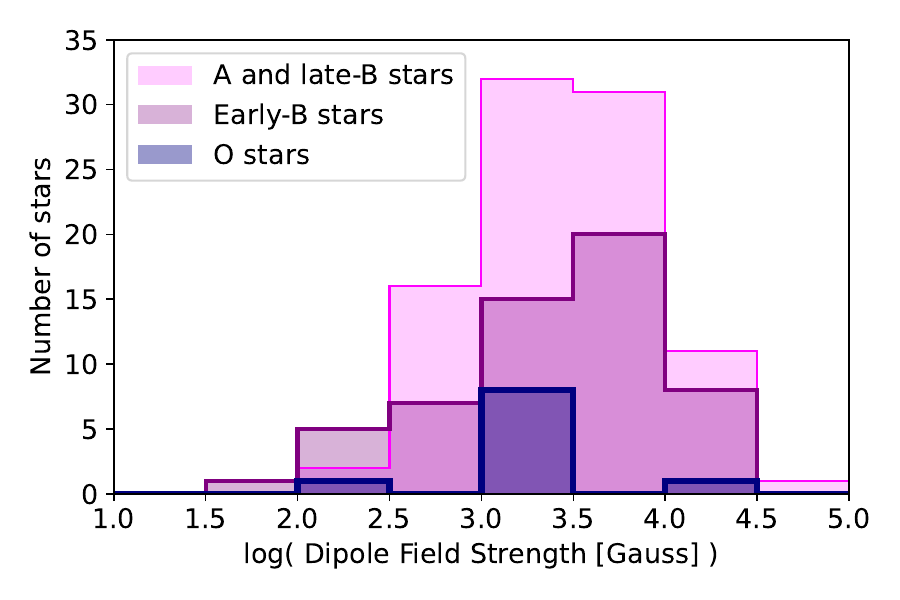}
\caption{Distribution of magnetic dipolar strengths for A and late-B stars, early-B stars, and O stars. Note that the x-axis is in $\log_{10}$. \\
Adapted from Fig. 6 of \citet{2019MNRAS.490..274S} -- small differences might be due to the specific columns used in the data sources, and the way the multi-pole results were tabulated (here we use the dipolar components only for simplicity): \citet{2007A&A...475.1053A} Table 2 column $B_d$,  \citet{2019MNRAS.483.3127S} Table 4 column $B_d$, \citet{2000A&A...359..213L} Table 1 column $B_d$ (absolute value), \citet{2019MNRAS.490..274S} Table C2 column $B_p$, 
\citet{2013MNRAS.429..398P} Table 1 column $B_p$ with the same modifications and additions as listed in \citet{2019MNRAS.490..274S}: 1000G for HD 108 from \citet{2017MNRAS.468.3985S}, 140G for $\zeta$ Ori A from \citet{2015A&A...582A.110B}, and the addition of HD 54879 with the 1920G measurement from \citet{2022MNRAS.510.4405J}.
}
\label{Bfield:fig:Bpole}
\end{figure}

Fig. \ref{Bfield:fig:Bpole} shows the distributions of magnetic field strengths (in kG) for a set of known O stars, early-B stars and late-B/A stars, from the sources listed in the caption. The distributions are very similar. \citet{2019MNRAS.490..274S} found that these distributions agree with a log-normal distribution with a mean of $10^{3.5}$ G and a standard deviation of $10^{0.6}$ G. Some of the stars with the largest magnetic fields known are in the range of tens of kG. 
Interestingly, if one was to shrink the most magnetic OB stars known to the size of a neutron star while conserving the magnetic flux (see Sect.~\ref{Bfield:sec:evolve}), the surface field would be in the same order of magnitude as those of magnetars with $\sim10^{15}$ G \citep{2008MNRAS.389L..66F}.

On the lower end of the distribution, there are very few known magnetic stars with magnetic fields less than a few hundred Gauss. This could have been due to a magnetic detection threshold that prevents us from finding stars with small fields. 
However, \citet{2007A&A...475.1053A} performed a survey of the Ap stars with the weakest polarization signals and found that they still have dipolar magnetic fields $>$300 G. Furthermore, surveys of non-Ap, A-type stars have not yielded any magnetic field detection with a precision below 300 G in most cases \citep[e.g.][]{2006A&A...450..777B}. This lack of ApBp stars with $\sim10$--100 G is called the \textit{Magnetic Desert}. 

This magnetic desert in ApBp stars was highlighted even more with the recent detection of so-called ultra-weak magnetic fields at the surface of the bright A-type stars Vega, Sirius A, $\beta$ UMa, $\theta$ Leo, and Alhena \citep{2009A&A...500L..41L, 2011A&A...532L..13P, 2016A&A...586A..97B, 2016MNRAS.459L..81B}. The shape of the circular polarization signal does not match what would be expected from the standard Zeeman effect theory for fossil fields. This said, if this signal is interpreted as a magnetic signal it would correspond to magnetic fields at the $\sim1$ G level. 
These ultra-weak magnetic stars are still rare, because these observations are at the limit of the current technological capabilities of spectropolarimetry. This means that perhaps weak magnetism may be more ubiquitous than previously thought. On the other-hand, these Stokes V detections also underscore the presence of the magnetic desert -- if $\sim$1 G level magnetic fields are detectable in A stars, magnetic fields of $\sim10$--100 would have been detected by now, if they existed in main-sequence A-type stars. 

For more massive O and early B-type stars, the detection capabilities are, in general, not as good as for A and late-B stars and highly depend on the characteristics and distance of the stars. 
For example, \citet{2019MNRAS.489.5669P} analyzed the dipolar field detection threshold for the O-type stars in the MiMeS Survey, and found that they ranged from 10 G to 1000 G, with a median of 85 G. From a Monte-Carlo analysis, if dipolar fields with strength of about 100G were ubiquitous in the MiMeS survey, they would have been found, which strengthens the idea that a magnetic desert might also exist for O-type stars. 

As can be seen in Fig. \ref{Bfield:fig:Bpole}, the distribution of magnetic B-type stars extend to slightly lower field strengths. Two B-type stars that are in the last third of the main sequence have been found with magnetic field dipolar strengths that could be lower than the strength threshold for ApBp stars \citep[$\beta$ CMa $B_d>60$ and $\epsilon$ CMa $B_d>13$G][]{2015A&A...574A..20F}. This could be indicative of a mass or age dependence on the Magnetic Desert's threshold.

\subsection{Topology of magnetic fields on the main sequence}

Large surveys of the $B_\ell$ variation of the magnetic field of OBA stars have shown that they mainly have geometries that are close to dipolar \citep[e.g.][]{2005A&A...430.1143B}. This said, when either precise $B_\ell$ curves are obtained or direct modeling of the Stokes V profiles is performed (e.g. ZDI), some additional medium-scale structures are nearly always found. The complexity of the magnetic field is often parametrized by either a multipole expansion (dipole$+$quadrupole$+$octople$+...$) or with spherical harmonics ($Y_\ell^m(\theta,\varphi)$). For example, Table 1 of \citet{2000A&A...359..213L} shows that many Ap stars require significant quadrupolar and octopolar components to reproduce their $B_\ell$ variations with rotation. ZDI modeling of ApBp stars and O/early-B stars have also shown that some higher harmonics are often necessary. At the extreme, HD 37776 (B2IV) and $\tau$ Sco (B0.2V) are quite famous for their field structure for which the $\ell= 3-4$ modes dominate \citep{2011ApJ...726...24K,2006MNRAS.370..629D}.

The tilt of the main dipolar component (the obliquity angle $\beta$ defined in Sect.~\ref{Bfield:sec:measure}) has been best studied for ApBp stars. \citet{1970ApJ...159.1001L} found that the magnetic axes are randomly distributed on the star, which means that larger $\beta$ will be favored with respect to the rotational axis. This is because the only one way to get $\beta=0$ is for the magnetic axis to be aligned with the rotational axis, but there are many more directions around a star that correspond to $\beta=90$. 

Surprisingly, however, \citet{2000A&A...359..213L} found that for Ap stars with \textit{long} rotational periods ($>25$d) the magnetic axes tend to align more closely with the rotational axis (small values of $\beta$). 
For magnetic early-B stars, \citet{2019MNRAS.490..274S} found a distribution of $\beta$ angles compatible with a random orientation of the magnetic axis.

\begin{BoxTypeA}[Bfield:interesting]{Did you know?}
\begin{itemize}
\item HD 215441 (Babcock's star) is the ApBp star with the strongest magnetic field \citep[34 kG,][]{1960ApJ...132..521B}.
\item NGC 1624-2 is the O-type star with the largest magnetic field \citep[20 kG,][]{2012MNRAS.425.1278W}.
\item HR 5907 is the magnetic early-B star with the fastest rotation period \citep[0.5 d,][]{2012MNRAS.419.1610G}. 
\item HD 60431 is the ApBp star with the fastest rotation period \citep[0.4 d,][]{2022A&A...668A.159M}.
\item HD 965 is the ApBp star with the longest period for which the magnetic field has been followed through a whole rotation \citep[16 yr,][]{2019A&A...629A..39M}.
\item $\tau$ Sco and HD 37776 are two magnetic B-type stars with un-usually complex magnetic fields. 
\end{itemize}
\end{BoxTypeA}

\subsection{Magnetic Stars in binaries on the main sequence}

Magnetic Ap stars are in wide binary systems as often as non-magnetic stars \citep{1973ApJS...25..137A}. However, there is a complete lack of Ap in binaries with orbital periods shorter than 3 days \citep{2002A&A...394..151C}. 
There are also plenty of magnetic O and early-B stars in wide binary systems. However, the \textit{Binarity and Magnetic Interactions in various classes of Stars} (BinaMIcS) Survey has only found a magnetic incidence of less than 2\% in close binary systems with periods less than 20 days \citep{2015IAUS..307..330A}.
Overall, there are only about a dozen of known magnetic massive stars binaries with $P_\mathrm{orb} < 20$ d \citep[see Table 5 of][]{2017A&A...601A.129L}.
There is only one known close binary system with two magnetic OBA stars: the B+B system $\epsilon$ Lup with an orbital period of 4 days \citep{2015MNRAS.454L...1S}. 

Although rare, these binary systems are very interesting. For example, such binary stars are expected to be co-evolved. It is therefore a challenge for the ISM formation scenario to explain why one star is magnetic, and the other not.  The lack of close binaries is certainly a point in favor of the merger field generation scenario. It is not clear, however, if this scenario can explain a system like $\epsilon$ Lup. 

\begin{BoxTypeA}[Bfield:binary]{Some noteworthy magnetic binary systems}
\begin{itemize}
\item Plaskett's star (HD 47129) is the only close massive binary that contains a magnetic O-type star \citep{2013MNRAS.428.1686G, 2022MNRAS.512.1944G}. It is also the only fast-rotating magnetic O-type star. 
\item HD 45166 is a binary system containing a low-mass (2$M_\odot$) Wolf-Rayet (WR) star with a magnetic field of 43 kG in a wide orbit around a non-magnetic late-B star. There is good evidence that the WR star might be a merger product \citep{2023Sci...381..761S}. HD 148937 is another potential merger product \citep{2024Sci...384..214F}.
\item HD 66051 is the first eclipsing binary that contains an early-type star \citep{2018MNRAS.478.1749K}.
\end{itemize}
\end{BoxTypeA}

\subsection{Evolution of magnetic fields in massive stars}
\label{Bfield:sec:evolve}

The magnetic field strength of a fossil field may change for a few reasons.  If there are no other mechanisms to increase or decrease the strength of a fossil magnetic field, it will experience a slow decline called Ohmic decay.  This decay has a characteristic time-scale that is given by
$$
t_{Ohmic} = \frac{4 \pi L^2 \sigma}{c^2},
$$
where $L$ is the scale size of the system, $\sigma$ is the electrical conductivity, and $c$ is the speed of light. If the magnetic field strength declines faster or slower than this time-scale, there may be other effects or stellar changes driving increases or decreases in the field strength.  As an important example for evolving stars, if a star experiences an increase in radius, that change in radius should affect the surface magnetic field because of the concept of flux conservation. Because most massive stars are mainly ionized, plasma flows cannot cross magnetic field lines due to the Lorentz forces. Therefore, either the plasma has to flow along field lines, or the field lines have to get dragged by the flow, depending on the relative strength of the field and the flow. In the case of the interior of a star, the latter generally applies. As a star expends in radius, the field lines will find themselves further apart, and the density of field lines (which is the definition of magnetic field strength) will be lowered. 

The magnetic flux over the surface of the star is generally approximated by $\phi \approx 4\pi R^2 B_{surf}$ (and often, the factor of $4 \pi$ is even ignored). Assuming an initial radius and surface magnetic field, for example the radius at the zero-age main sequence, the effect of a radius change on the surface magnetic field strength under flux conservation can be computed from

$$
B_{surf}(t) =  B_{surf}(t_{ZAMS}) \left(\frac{R(t_{ZAMS})}{R(t)}\right)^2 .
$$

This relationship implies that as the star increases in radius, its surface magnetic field would be expected to decrease as the square of the radius.  Additionally, the development of convective outer layers that may develop during post-main sequence evolution \citep[e.g.,][]{2009ApJ...705.1000F} could complicate the large-scale magnetic fields of OBA stars, creating the possibility of dynamo fields developing in those layers, either enhancing or diminishing any surface fields.  

\subsubsection{Magnetic evolution on the Main Sequence}
Observationally, we find a change in field strength of OBA stars as a function of stellar age -- this is expected, as the radius is expected to change by a factor of a few over the MS \citep{2008A&A...481..465L,2019MNRAS.490..274S}. 
An interesting question is whether the magnetic flux is observationally conserved on the MS. To answer this, we need to determine the expected change in radius from evolution models. Typically, the associated additional systematic uncertainties are such that the age determination needs to be rather precise to evaluate the evolution of the magnetic flux. The best age estimates come from stars that are members of clusters. This is especially true for ApBp stars, because the uncertain bolometric correction calibration for ApBp stars \citep[due to their abundance peculiarities impacting their color, as discussed in detail in][]{2007A&A...470..685L} makes their position on the HRD more uncertain.  

\citet{2008A&A...481..465L} found that for ApBp cluster members with masses between 2$M_\odot$ and 5$M_\odot$, the observations are constant with a small decrease in magnetic flux over the MS lifetime. 
\citet{2019MNRAS.490..274S} found that the magnetic flux of field and cluster magnetic OB-type stars also seem to decrease on the main sequence by a factor of a few. Although they note that for the most massive stars, the evolution tracks used to determine stellar ages and main sequence lifetimes do not include the impact of magnetism on stellar structure and evolution, which might be significant for the most massive O-stars.

Another approach is to study the incidence of magnetic stars as a function of time. For example, \citet{2016A&A...592A..84F} constructed a probability density distribution of fractional main-sequence age for all of the magnetic OB stars known at the time. They compared this distribution to that of a representative magnitude-limited survey of normal OB stars and the expected sensitivity from modern spectropolarimetric instruments. They found that the age distribution of the comparison sample was dominated by stars in the last quarter of their MS lifetime, whereas the distribution of magnetic stars peaked in the middle of the main sequence lifetime. This would indicate a lack of old magnetic MS stars. Unfortunately, the detection limits actually achieved by the MiMeS Survey — at least for O-type stars, while compatible with magnetic flux conservation, were not low enough to convincingly establish constraints on flux decay on the MS \citep{2019MNRAS.489.5669P}. 

\subsubsection{Magnetic evolution after the Main Sequence}

As stars leave the main sequence, ceasing hydrogen fusion reactions in their core,  they will typically experience physical changes to surface temperature, physical radius, and some may experience a change in structure, i.e., the development of a convective layer.  In these more evolved stars, the large-scale stable magnetic fields seen in main sequence OBA stars will inevitably be affected by such changes.   New modeling and theoretical studies have aimed to build in the possible effects that these fields may have on the interior structure and evolutionary path of OBA stars \citep[see e.g.,][]{2003A&A...411..543M, 2016A&A...585A.120S}.  Many studies and more recent work, for example the work of \citet{2019MNRAS.485.5843K} (and other papers in that series), have tried to build in the effects of the magnetic field in a physically realistic manner.  For more information on this, we recommend the recent review article by \citet{2023Galax..11...40K}.  Understanding how the magnetic field affects the star as it evolves and vice versa remains a challenging area of research.

Efforts to understand the problem from the observational side have faced different challenges.  One of the first confirmed magnetic fields in a star evolved from an OBA star was EK Eridani and the study of \citet{2008A&A...491..499A}.  The first O supergiant with a weak detected magnetic field was $\zeta$ Ori Aa, which was determined to be magnetic by \citet{2008MNRAS.389...75B}, and confirmed by \citet{2015A&A...582A.110B}. Several surveys have investigated red giant and supergiants that have evolved from main sequence OBA stars \citep[see e.g.,][]{2010MNRAS.408.2290G,2015A&A...574A..90A}, and found for these stars that dynamo fields did develop in convective zones, affecting the observation of any previous main sequence magnetism due to a fossil field on the stellar surface.  

The discovery of two significantly evolved A supergiants, $\iota$ Car and HR 3890, with longitudinal magnetic field strengths of just a few G \citep{2017MNRAS.471.1926N}, confirmed that study of evolved OBA stars would require significantly lower uncertainties due to small field strengths that were consistent with flux conservation.  From this the \textit{Large Impact of magnetic Fields on the Evolution of hot stars} (LIFE) project was started to systematically use high-precision spectropolarimetric observations to investigate the magnetism of OBA supergiants.  The goal was to understand how stellar magnetism works in their transitional evolutionary phases.   \citet{2018MNRAS.475.1521M} reported early results, which appear to be consistent with magnetic flux conservation, but the overall survey results are still in progress.  This type of study faces challenges: (1) high-precision observations of many supergiants require significant observational time, and (2) the number of such realistically observable targets is smaller, making significant statistical conclusions, such as those done for main sequence populations, more difficult.

\section{Conclusions}

Stellar magnetism affects all types of stars throughout their lifetimes.  
It can greatly modify their internal and external properties and how their evolutionary path proceeds.  
For OBA type stars, we find a fossil magnetic field different from that found for the Sun, and only a $\sim$10\% incidence rate.   We have discussed some of the known properties of these types of stars and some of the ways we observe and study them.

But, stellar magnetism is just one aspect of the physical processes involved in stars, therefore we cannot always consider it as acting on its own. 
We have mentioned throughout this chapter some cases of this as magnetic fields affect the atmospheres of stars, creating chemically peculiar stars with inhomogenous surface abundances, and the interaction of the magnetic field with stellar winds to create magnetospheric structures.  
Another intersection that stars with magnetic fields may have is with stellar pulsation (see the Chapter on Asteroseismology in this Encyclopedia). 

Several categories of OBA pulsating stars are also known to possess a fossil magnetic field.  roAp stars have been long known to have strong dipolar magnetic fields and stellar pulsations \citep{1990ARA&A..28..607K}.  
Both \textit{slowly pulsating B} (SPB) stars \citep[e.g.,][]{2003A&A...406.1019N} and $\beta$ Cep \citep[see e..g.,][]{2013A&A...555A..46H, 2012MNRAS.427..483B} stars have measured magnetic fields that have been well studied also in pulsation observations \citep[see e.g.,][]{2016A&A...587A.126B}. 
Most recently, the pulsating $\delta$ Scuti stars have been discovered to have magnetic fields \citep[see e.g,][]{2015MNRAS.454L..86N}.

Studying the interiors of stars using asteroseismology is already a challenging task, but the presence of a strong fossil magnetic field can affect various aspects of the pulsational observations.  For example, pulsational frequencies are split by the presence of the magnetic field.   
The properties of the resulting frequencies depends on both the strength of the magnetic field and the magnetic geometry \citep[see e.g.,][]{2000ApJ...531L.143S}.  
The magnetic field can also affect the period spacing for gravity-mode pulsations, which can then be used to estimate the strength of the internal magnetic field \citep[see e.g.,][]{2020A&A...636A.100P}. Recent work by \citet{2022Natur.610...43L} has used this effect to infer the strength of the internal magnetic fields in the cores of a sample of red giants.

To properly study the dual effects of pulsation and magnetic fields in a star requires a new type of technique.  Magneto-asteroseismology evaluates the magnetic effects and the pulsational effects simultaneously to accurately determine both properties.  
This technique is crucial to understand the impact of magnetism on the internal dynamics and evolution of these stars.  
Thus far, only one star, the SPB HD 43317 has been analyzed combining both magnetic and pulsational effects \citep{2018A&A...616A.148B}, and by adding in MHD simulations, \citet{2022MNRAS.512L..16L} were able to make the first estimate of the magnetic field strength inside a star on the main sequence.  
Stellar asteroseismology and magneto-asteroseismology are a relatively new field and technique that will grow as high-precision, long-term photometry continues to become available.

Magneto-Asteroseisomology is just one example of studying two effects together to gain a more complete picture of the effects of physical processes.  We can imagine that the future of studying magnetic fields in stars will continue to connect with other physical mechanisms as we learn more about how they are intertwined.  

\begin{ack}[Acknowledgments]

VP is supported by the National Science Foundation (NSF) under Grant AST-2108455. MEO is supported by the National Science Foundation (NSF) under Grant AST-2107871. We thank the editor Fabian Schneider for expert feedback and insightful suggestions. 

\end{ack}

\seealso{\citet{2009ARA&A..47..333D}, \citet{2015SSRv..191...27L}}

\bibliographystyle{Harvard}
\bibliography{reference}

\end{document}